\documentclass[review]{elsarticle}

\usepackage{lineno,hyperref}
\modulolinenumbers[5]

\journal{Journal of \LaTeX\ Templates}

\usepackage{amssymb}
\usepackage{color}

\begin{document}

\begin{frontmatter}

\title{Advances in the optimization of silicon-based thermoelectrics: a theory perspective}
%\tnotetext[mytitlenote]{Fully documented templates are available in the elsarticle package on \href{http://www.ctan.org/tex-archive/macros/latex/contrib/elsarticle}{CTAN}.}

%% Group authors per affiliation:
\author{Davide Donadio}%\fnref{myfootnote}}
\address{Department of Chemistry, University of California Davis, One Shields Ave. Davis, CA, 95618, USA}
%\fntext[myfootnote]{Since 1880.}

%% or include affiliations in footnotes:
%\author[mymainaddress,mysecondaryaddress]{Elsevier Inc}
%\ead[url]{www.elsevier.com}

%\author[mysecondaryaddress]{Global Customer Service\corref{mycorrespondingauthor}}
%\cortext[mycorrespondingauthor]{Corresponding author}
\ead{ddonadio@ucdavis.edu}

%\address[mymainaddress]{1600 John F Kennedy Boulevard, Philadelphia}
%\address[mysecondaryaddress]{360 Park Avenue South, New York}

\begin{abstract}
Thermoelectric devices convert temperature gradients into electrical power and {\it vice versa}, thus enabling energy scavenging from waste heat, sensing and cooling.
Yet, many of these attractive applications are hindered by the limited efficiency of thermoelectric materials, especially in the low temperature regime.
This review provides a summary of the recent advances in the design of new efficient silicon--based thermoelectric materials through nanostructuring, alloying and chemical optimization, emphasizing the contribution from theory and atomistic modeling. 
\end{abstract}


%Declarations of interest: none

\end{frontmatter}

%\linenumbers

%Measuring and harnessing heat is essential in large variety of technologies, including microelectronics, high-resolution imaging, sensing and renewable energy harvesting.
Thermoelectric (TE) devices can be used to scavenge thermal energy, and can be combined to photovoltaics to enhance energy conversion efficiency. Peltier coolers will play an increasingly important role in active heat management at the small scale, for example in electronic devices and sensors. However, the large-scale implementation of TE technology stems from improving the efficiency and the processing costs of materials. 
The conversion efficiency of TE materials is determined by their dimensionless figure of merit, 
\begin{equation}
ZT = \frac{\sigma S^2 T}{\kappa_e + \kappa_p}, 
\end{equation}
where $\sigma$ is the electrical conductivity, $S$ the Seebeck coefficient, and $\kappa_e$ and $\kappa_p$ the electronic and phononic contributions to the thermal conductivity. The electronic coefficients, $\sigma$, $S$ and $\kappa_e$ are tightly intertwined and can hardly be optimized separately, whereas the lattice thermal conductivity is decoupled from the other properties. This observation lead to the suggestion that high $ZT$ may be achieved in materials that behave like "electron crystals and phonon glasses" \cite{Nolas:1999kv,He:2017cn}. However, this condition occurs in few families of materials, such as skutterudites,\cite{Nolas:1999kv,RullBravo:2015cu} clathrates,\cite{Takabatake:2014cu}, Zintl compounds  and a number of heavy metal chalcogenides.\cite{Snyder:2008jh,Alam:2013ct} 

\begin{figure}[htbp] %  figure placement: here, top, bottom, or page
   \centering
   \includegraphics[width=4in]{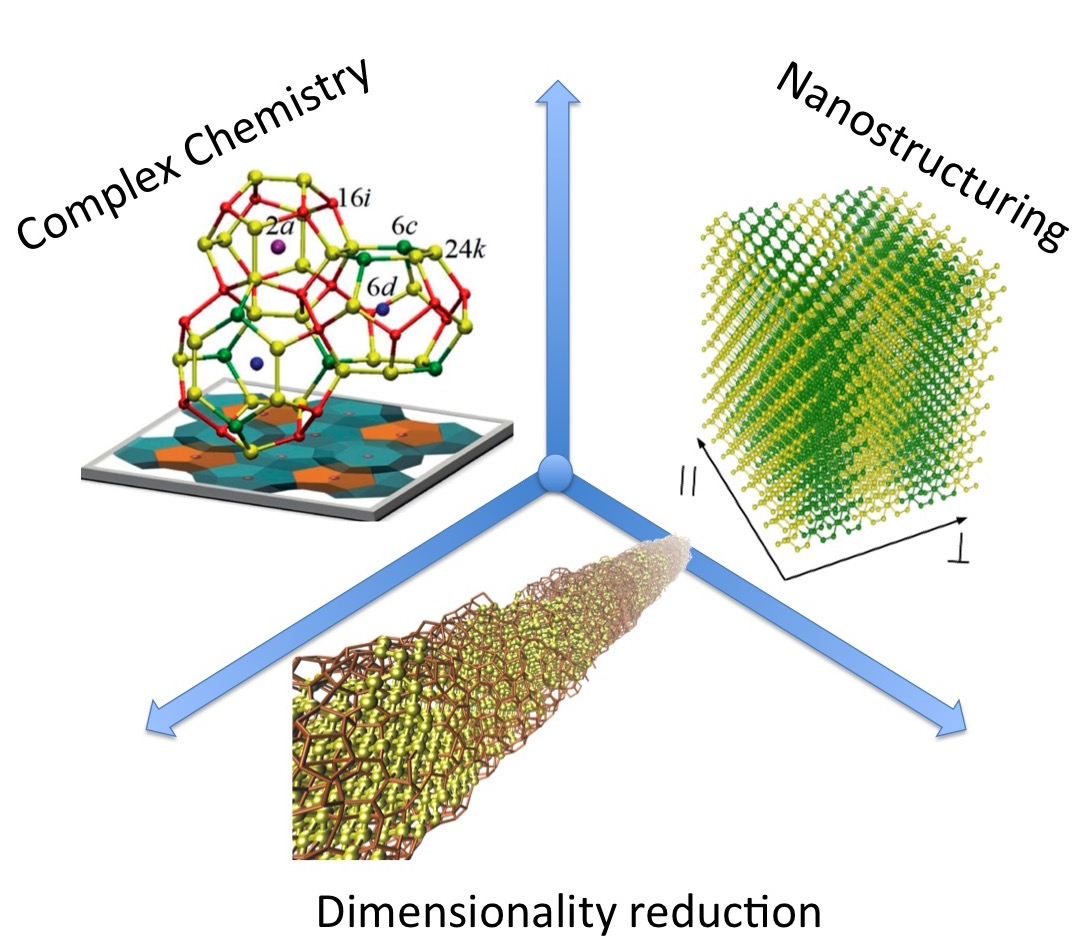} 
   \caption{Possible design routes to optimize silicon based thermoelectric materials involve a multidimensional parameter space that comprises chemical complexity, nanostructuring and dimensionality reduction. Examples of outcomes of these optimization strategies are silicon-based clathrates, SiGe superlattices and silicon nanowires with rough surfaces. Models are adapted from Refs. \cite{Kauzlarich:2016el,Savic:2013tg,Donadio:2010kp}.} 
   \label{Fig:opti}
\end{figure}

%Promising calculations on model systems stimulated experimental research on nanostructured thermoelectric materials that lead to the breakthrough discovery of materials with $ZT>1$ \cite{Majumdar:2004fm,Venkatasubramanian:2001iy,Poudel:2008cj}. More recently, hierarchical nanostructuring lead to record high $ZT$ of 2.4 for PbTe alloyed with SrTe and doped with Na\cite{Biswas:2012fw}. 
Most of these materials are either toxic or expensive, as they contain rare elements, and doping cannot be easily controlled.  
Conversely, the use of silicon as TE material would be ideal, as it is abundant, non-toxic and easy to process and to dope, both $p$ and $n$, but bulk crystalline silicon is a rather poor thermoelectric ($ZT~0.01$ at room temperature).  
In spite of suitable electronic conductivity and relatively high Seebeck coefficient, the TE efficiency of crystalline silicon is mostly hampered by high lattice thermal conductivity ($\kappa_p=148 - 156$ Wm$^{-1}$K$^{-1}$ at 300K for natural and isotopically pure samples, respectively \cite{Kremer:2004dv}). 
Alloying was readily identified as a practical way to reducing $\kappa_p$ while retaining a crystalline structure and a viable thermoelectric power factor ($\sigma S^2$)\cite{doi:10.1063/1.1722984}. SiGe alloy has indeed been used for high temperature TE energy generation since the 1960s for niche applications, such as powering space probes, \textcolor {black}{and a record high $ZT=1.5$ was achieved at $\sim1250$ K \cite{Bathula:2012iy}}. Yet, the deployment of TE silicon at low-temperature requires further optimization, either by enhancing the power factor through band engineering, or by reducing $\kappa_p$ \cite{Neophytou:2015tw}. These two goals may be achieved through different strategies, by moving in a multidimensional optimization space that encompasses chemical complexity, nanostructuring and dimensionality reduction (Figure~\ref{Fig:opti}). 
Hereafter recent progress in the optimization of silicon-based TE are reviewed, focusing on the contribution from theory and modeling.

\section{Electronic Band engineering.} 

The electrical transport coefficients that appear in the expression of $ZT$ are tightly intertwined: at fixed charge mobility $S$ declines as a function of the density of charge carriers while $\sigma$ increases. In addition, $\sigma$ is proportional to $\kappa_e$ through Wiedmann-Franz law.  While these fundamental relations limit the possibilities to improve $ZT$ through band engineering, it was observed that Seebeck coefficient depends on the derivative of the electronic density of states (eDOS) at the Fermi level. Hence, a very narrow, ideally delta-shaped, charge carriers distribution would maximize $\sigma S^2$\cite{Mahan:1996te}. In the early 1990s Hicks and Dresselhaus suggested that nanostructuring \cite{Hicks:1993gh} and dimensionality reduction \cite{HICKS:1993p3198} can largely enhance $S$ as a result of quantum confinement. This effect may be achieved either by  incorporating quantum wells, quantum wires or quantum dots in bulk silicon,\cite{Dresselhaus:2007hx} or by growing low-dimensional nanostructures. Measurements on silicon nanowires \cite{Boukai:2008iz} and on highly-doped silicon thin films \cite{Ikeda:2010fb} corroborate these earliest predictions. 

It is indeed possible to modify the band structure by quantum confinement in low-dimensional systems: density functional theory (DFT) and \textcolor {black}{many-body perturbation theory GW calculations} show that the energy gap of nanometer-thin silicon wires and membranes is inversely proportional to their thickness and undergoes a transition from indirect to direct below a threshold thickness.\cite{Vo:2006kd,PhysRevB.76.113303,Mangold:2016gw} 
However, these substantial changes in the electronic structure of silicon nanostructures do not cause an enhancement of $S$, unless one employs extremely thin wires or membranes, with characteristic diameter/thickness $<2$ nm \cite{Vo:2008kx,Markussen:2009fz,Fagas:2009go,Mangold:2016gw}. \textcolor {black}{Below this threshold sharp features appear in the eDOS that enhance $S$, but $\sigma$ is substantially reduced.} The overall enhancement of the TE power factor ($\sigma S^2$) is then negligible.  
Calculations of transport coefficients based on the semi-classical Boltzmann transport equation (BTE) unravel the concurrent reasons for this result. 
First of all, enhancement of $S$ over the bulk value was predicted only for the thinnest wires and membranes, because the eDOS does not exhibit sharp peaks at the band edge unless confinement is of the order of 1 or 2 nm. Secondly, although at extreme confinement one may expect a potential enhancement of mobility, up to a factor 4 \cite{Vo:2008kx}, the modified eDOS leads to a considerable reduction of the overall electrical conductivity, without considering the effect of surface defects, which may hamper $\sigma$ even further\cite{Evans:2005dt}. For nanostructures with more viable characteristic size, i.e. thickness $\sim 10$ nm, both DFT and tight-binding calculations predict charge transport coefficients close to those of bulk silicon.\cite{Neophytou:2013ee,Neophytou:2015tw,Mangold:2016gw}

Remarkably, standard band transport models cannot justify the increase of $S$ measured in Refs.\cite{Boukai:2008iz,Ikeda:2010fb}. So high values of $S$ at low temperature in highly doped low-dimensional materials may stem from phonon drag, \textcolor {black}{which is an increase to the effective mass of charge carriers induced by the interaction of electrons with phonons}. A first-principles theory of phonon drag was recently developed and showcased on bulk silicon.\cite{Zhou:2015jo} This approach requires a well-converged sampling of electron-phonon interactions over the Brillouin Zone and it has not yet been applied to complex systems, but it opens the way to future optimization perspectives.

Nanostructured silicon-based materials, such as nanoporous and nano-crystalline silicon were also investigated as potential TE materials and their charge transport coefficients were computed. DFT calculations indicate that the power factor of nanoporous silicon (np-Si) is only slightly reduced compared to that of bulk Si, even at high porosity\cite{Lee:2008tn}. 
For boron-doped nanocrystalline silicon $S$ may increase beyond that of bulk because of both carrier filtering and temperature discontinuities at grain boundaries. Concurrently also the electrical conductivity may be increased due to the higher Fermi level of grains compared to bulk Si at the same concentration of holes. The combination of these effects results in up to 5 times higher power factor than in the bulk.\cite{Neophytou:2013iw,Narducci:2014fi} These calculations may account for the high $ZT$ formerly measured for B-doped Si/SiGe composites \cite{Zebarjadi:2011ge}.   

More aggressive band structure engineering may be achieved through the synthesis of metastable phases with higher chemical complexity. For example, experimental efforts devoted to the optimization of silicon-based clathrates, by exploring their compositional space, and achieved $ZT~0.5$ at 1200 K\cite{Kauzlarich:2016el}. High $ZT$ is favored by extremely low values of $\kappa_p<1$ $Wm^{-1}K^{-1}$,\cite{He:2014ge} but the power factor is hampered by high electrical resistivity.\cite{Sui:2016ki} Although both the electronic structure and the phonons of silicon-based clathrates have been calculated for their potential as superconductors,\cite{Connetable:2007bh} a systematic modeling-driven effort toward the TE optimization of this class of materials is still lacking.

%%%%%%%%%%%%%%%%%%%%%%%%%%%%%%%%%%%%%%%%%%%%%%%%%%%%%% 
\section{Phonon Transport Engineering} 

\textcolor {black}{The calculations and the measurements discussed in the previous section suggest that silicon-based materials may not achieve high  $ZT$ through the improvement of $\sigma S^2$ alone, unless $\kappa_p$ is reduced by about a factor 100 with respect to the bulk value.}
Major efforts have been deployed to understand phonon transport and engineer low-$\kappa_p$ materials.
From the solution of the phonon Boltzmann transport equation (PBTE), the contribution of each phonon branch $i$ to $\kappa_p$ is expressed as:
 \begin{equation}
 \kappa_i(q) = c_i(q) \mathrm{v}_i(q) \lambda_i(q)
 \label{eq:kappa}
 \end{equation}
 which has to be integrated over the wave vector $q$ in the Brillouin zone. $c_i$ is the mode heat capacity, v$_i$ the group velocity component in the direction of propagation and $ \lambda_i$ the phonon mean free path (mfp).\cite{ziman} The latter can be written as the product of group velocity and relaxation time ($\tau_i$).
% : $\lambda_i=\mathrm{v}_i \tau_i$. 
 $\tau_i$ is dictated by anharmonic phonon-phonon scattering, defects scattering and surface or boundary scattering. In earlier works scattering rates were parametrized on experimental data, and  $\lambda$ was determined empirically.\cite{Callaway:1959uo,Holland:1963uo} More recently it has been possible to calculate $\tau_i$ and therefore $\kappa_p$, solving the linearized PBTE and computing the interatomic force constants by first-principles including three- and four-phonon scattering terms.\cite{Broido:2007iu,Fugallo:2013bm,Cepellotti:2016bk,Ravichandran:2018cj}
Eq.~\ref{eq:kappa} suggests that there are two ways to engineer phonon heat transport: one is to modify phonon dispersion relations, which impact the density of states, the group velocities and the space of viable anharmonic scattering processes. The other is to modify the extrinsic scattering processes that contribute to define phonon lifetimes. %The former is more effective than the latter(?)  
%Dimensionality reduction and nanostructuring were readily recognized to have a stark effect on heat transport.\cite{Dresselhaus:2007hx,Joshi:2008dd} 

\subsection{Low-Dimensional Silicon}
\textcolor {black}{Dimensionality reduction and nanostructuring have a stark effect on phonons and heat transport.}
Pioneering measurements on crystalline silicon thin films~\cite{Asheghi:1997ke} and nanowires~\cite{Li:2003bo} showed significant reduction of $\kappa_p$ for systems with characteristic thickness of several tens of nm. The reduction of $\kappa_p$ is greatly enhanced by surface roughness, thus making rough silicon nanostructures viable TE materials.\cite{Hochbaum:2008hl,Chen:2008ig} More recently dramatic reduction of $\kappa_p$ was observed in ultra-thin silicon membranes, with thickness below 10 nm.\cite{ChavezAngel:2014be}
This effect was mostly ascribed to surface scattering, and kinetic models  reproduced well experimental measurements, including temperature trends.\cite{Mingo:2003hu}  In particular the effect of boundary scattering in  nanowires and thin films was modeled according to Casimir limit, i.e. assuming that the diameter or thickness are an upper boundary for $\lambda$. According to this model intrinsic bulk mfp ($\lambda_{bulk}$) are modified following Matthiessen's rule, so that the mfp of a nanostructure with characteristic size $D$ is $1/\lambda_i(\omega)=1/\lambda_{bulk}(\omega) + 1/D$, where $\lambda$ is expressed in terms of frequency ($\omega$) instead of wave vector $q$.
This approximation fails when $D \lesssim 20 nm$, because (i) parametrized kinetic models neglect the effect of dimensionality reduction on phonon dispersion relations, and (ii) empirical approaches such as Casimir limit or diffuse scattering models do not take into account the fine-grained details of surface effects, such as roughness,\cite{Martin:2009hl} chemical composition and resonances.    

As for (i) lattice  dynamics calculations and Brillouin scattering measurements show that in thin membranes, the dispersion relations of acoustic transverse acoustic mode with out-of-plane polarization (ZA modes) turns from linear to quadratic. The thinner the membranes, the softer the ZA modes.\cite{Cuffe2012,Neogi:2015cp,Neogi:2015gk} Similarly, in nanowires there are two  ZA modes with quadratic dispersion and higher order quadratic optical branches.\cite{Thonhauser:2004db,Donadio:2009fo} Dimensionality reduction alone is responsible for a mild decrease of $\kappa_p$ in nanostructures. $\kappa_p$ of ideally crystalline silicon membranes decreases to about $0.35\kappa_{bulk}$ for thickness of about 1 nm.\cite{Neogi:2015cp} For silicon nanowires the relation between $\kappa_p$ and the diameter is non-monotonic:\cite{Ponomareva:2007fj}: the minimum $\kappa_p$ occurs for $D\sim3-4$ nm and it is $0.2\kappa_{bulk}$.\cite{Donadio:2010kp}

As for (ii) surface roughness and the presence of an amorphous surface layer, be it amorphous silicon or native silicon oxide, produce a much more substantial reduction of thermal conductivity, as observed in several atomistic simulations of nanowires and suspended membranes.\cite{Donadio:2009fo,Donadio:2010kp,Aksamija:2010cg,He:2012dn,Duchemin:2012ii,Neogi:2015gk} The effect of surface layers depends on the surface-to-volume ratio, which is determined by the diameter/thickness of nanowires/membranes.  
Rough surface oxide layers reduce $\kappa_p$ up to 40 times in silicon nanowires with 15 nm diameter, and up to 25 times in 10 nm thick membranes. 
The stark effect of native oxide layers on $\kappa_p$ was confirmed experimentally by Raman thermometry of ultra-thin silicon membranes before and after etching of the surface oxide layers.\cite{Neogi:2015gk}  
Recent studies revealed the microscopic mechanism that leads to $\kappa_p$ reduction. Contrary to the assumption of diffusive phonon scattering\cite{Maznev:2015di}, spectral energy decomposition~\cite{Thomas:2010do} and frequency--resolved calculations of phonon mfps\cite{Saaskilahti:2014bk} show that surface layers of amorphous materials mainly produce resonances that modify the dispersion relations of low-frequency propagating modes, reducing their group velocity.\cite{Zushi:2015fo,Xiong:2017if} 
These resonances are similar to those obtained by \textcolor {black}{nano-patterning} the surface of thin films, membranes or nanorods according to the recently proposed nanophononic metamaterials paradigm.\cite{Davis:2014hu,Honarvar:2016jo,Honarvar:2016dm,Xiong:2016bn,Anufriev:2017br}
 
Such reduction of $\kappa_p$ is sufficient to utilize silicon membranes for sensing \cite{Varpula:2017ds}, but not for TE energy conversion, as the maximum $ZT$ that can be achieved is 0.2 at room temperature \cite{Mangold:2016gw}. To achieve $ZT$ beyond 1 it would be necessary to further decrease $\kappa_p$. This can be obtained by introducing ``bulk" defects. Alloying with germanium, nanopores, twin boundaries, dislocations and heterostructures have been proposed as effective ways to further reduce $\kappa_p$,\cite{Chan:2010ba,Hu:2011bla,Hu:2012fe,He:2012dn,XiongPRB2014,Xiong:2017if} but their impact on charge transport coefficients still needs to be assessed. In particular, MD predicts that introducing 10$\%$ of Ge would decrease $\kappa_p$ for 10 nm membranes ten times leading to $ZT>1$.\cite{Xiong:2017if} 

\subsection{Nanoporous Silicon and SiGe Heterostructures}
Nanostructuring of bulk silicon is also an effective way to reduce $\kappa_p$. Based on early predictions from MD simulations,\cite{Lee:2007fr} nanoporous silicon (np-Si) films with extremely low $\kappa_p$ were manufactured and proposed as promising TE material~\cite{Tang:2010kv,Yu:2010fp}, but so far only at high temperature.\cite{deBoor:2012ea} A limiting factor is that the size and the spacing of the pores in experiments is usually one order of magnitude larger than that studied in atomistic simulations and the $\kappa_p$ reduction is not as drastic.\cite{Jain:2013cv,Jean:2014gw,Lee:2016jz} 
In np-Si $\kappa_p$ reduction originates from modified phonon dispersion relations, and the geometry and size of the pores can be designed to get $\kappa_p<1$ Wm$^{-1}$K$^{-1}$ at reasonably low porosity (30$\%$).\cite{Yang:2014co}
Similarly to what observed for nanostructures, a layer of amorphous material at the pore surface enhances the reduction of $\kappa_p$.\cite{He:2011fh,Verdier:2017fd}. Simulations predict that $\kappa_p$ reduction may be achieved by alloying with germanium \cite{Bera:2010dk,He:2011ig}

Heterostructures, such as silicon/germanium superlattices readily implement phononic band structure engineering.\cite{Hepplestone:2008js,Savic:2013tg} The reduction of $\kappa_p$ depends on the dimensionality of the inclusions and on superlattice spacing $L$. The relation between $\kappa_p$ and $L$ is non-monotonic and the minimum $\kappa_p$ depends on the dimensionality of the superlattice. This phenomenon, accounted for by atomistic simulations\cite{Garg:2011kt,Savic:2013tg} and verified in experiments,\cite{Pernot:2010fc,Luckyanova:2012jb} can be interpreted as a transition between a small-$L$ regime, in which Si/Ge interfaces produce phonon interference, and a large-$L$ regime, in which thermal boundary resistances add up to provide the total device resistance. The minimum $\kappa_p$ occurs at the highest density of interfaces before phonon interference kicks in. 
A promising approach to further the performance of SiGe superlattices is to grow films with concentration graded profiles \cite{FerrandoVillalba:2015hl,Hahn:2016hy}.   

\begin{figure}[htbp] %  figure placement: here, top, bottom, or page
   \centering
   \includegraphics[width=4in]{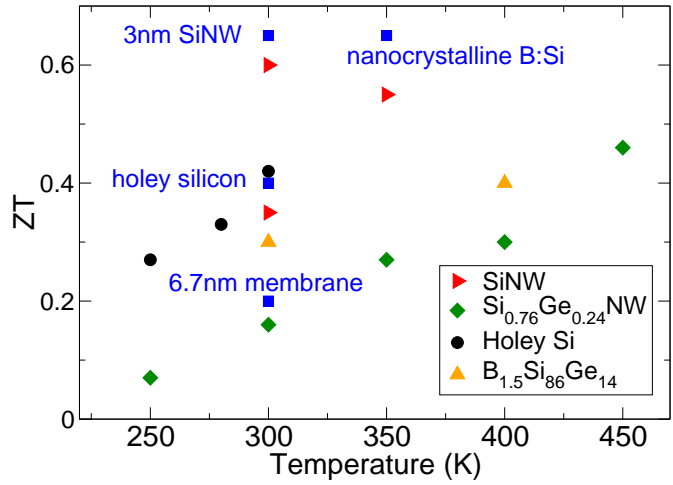} 
   \caption{Low-temperature thermoelectric figure of merit  ($ZT$) of low-dimensional and nanostructured silicon-based materials from theory and experiments as a function of temperature. Experimental measurements (see legend) are shown for silicon nanowires\cite{Boukai:2008iz,Hochbaum:2008hl}, SiGe nanowires\cite{Lee:2012fm}, nanoporous (holey) silicon\cite{Tang:2010kv} and B-doped nano crystalline SiGe\cite{Zebarjadi:2011ge}, and they are compared to theoretical predictions (blue squares and labels) for similar systems: SiNW \cite{Vo:2008kx}, membranes \cite{Mangold:2016gw}, nanoporous silicon \cite{Lee:2008tn} and B-doped nano crystalline Si \cite{Neophytou:2013iw}.} 
   \label{Fig:ZT}
\end{figure}

\section{Conclusions} 

A decade of optimization of the TE figure of merit of silicon-based materials has lead to $ZT$ significantly higher than that of bulk silicon, but has not yet reached the threshold for technological deployment of $ZT=1$ (see Figure~\ref{Fig:ZT}) Predictive theory and simulations have often preceded experiments and have provided a deeper understanding of charge and heat transport, unraveling new phenomena, such as the phonon resonance effect, and fostering the design of new nanomaterials, e.g. crystalline nanoporous silicon and ultra-thin silicon membranes.  
Simulations predict that it may be possible to achieve even higher $ZT$ by combining compatible optimization routes, e.g. alloying with nanostructuring \cite{Xiong:2017if,Xiong:2016bn,Lee:2012fm} or nano-crystallinity with nanoporosity \cite{Lorenzi:2018fw}.
However, several design concepts proposed by modeling still wait for experimental verification.
Finally, further efforts are required to achieve systematic materials design across the optimization space, possibly devising and applying high-throughput and machine-learning approaches that include search in the chemical space for complex nanostructured systems.\cite{Curtarolo:2013fa}

%In these works, however, it was eventually found out that rather than charge transport nanostructuring impacts thermal transport. 
%Thermal conductivity optimization: The idea of phonon band structure optimization. Early Exploratory works on SiGe composites: \cite{Dresselhaus:2007hx}

% Clathrates have been identified as good candidates TE materials, due to their low thermal conductivity.   
% Nanostructured silicon (superlattices and porous Si)

\section*{References}

%\bibliography{BigLib}

\end{document}